\documentclass[aps,prl,twocolumn,showpacs,amsmath,amssymb,superscriptaddress]{revtex4-1}
\usepackage{graphicx} 
\usepackage{epsfig}
\usepackage{dcolumn} 
\usepackage{color}

\begin{document}
\title{A topological quantum pump in serpentine-shaped semiconducting narrow channels}
\author{Sudhakar Pandey}
\affiliation{Institute for Theoretical Solid State Physics, IFW-Dresden, Helmholtzstr. 20, D-01069 Dresden, Germany}
\author{Niccol\'o Scopigno}
\affiliation{Institute for Theoretical Physics, Center for Extreme Matter and Emergent Phenomena, Utrecht University, Princetonplein 5, 3584 CC Utrecht,  Netherlands}
\author{Paola Gentile}
\affiliation{CNR-SPIN, I-84084 Fisciano (Salerno), Italy}
\affiliation{Dipartimento di Fisica ``E. R. Caianiello'',
Universit\`a di Salerno I-84084 Fisciano (Salerno), Italy }
\author{Mario Cuoco}
\affiliation{CNR-SPIN, I-84084 Fisciano (Salerno), Italy}
\affiliation{Dipartimento di Fisica ``E. R. Caianiello'',
Universit\`a di Salerno I-84084 Fisciano (Salerno), Italy }
\author{Carmine Ortix}
\affiliation{Institute for Theoretical Physics, Center for Extreme Matter and Emergent Phenomena, Utrecht University, Princetonplein 5, 3584 CC Utrecht,  Netherlands}
\affiliation{Dipartimento di Fisica ``E. R. Caianiello'',
Universit\`a di Salerno I-84084 Fisciano (Salerno), Italy }
\affiliation{Institute for Theoretical Solid State Physics, IFW-Dresden, Helmholtzstr. 20, D-01069 Dresden, Germany}

\begin{abstract}
We propose and analyze theoretically a one-dimensional solid-state electronic setup that operates as a topological charge pump in the complete absence of superimposed oscillating local voltages. The system consists of a 
semiconducting narrow channel
with strong Rashba spin-orbit interaction patterned in a mesoscale serpentine shape. 
A rotating planar magnetic field serves as the external ac perturbation, and cooperates with the Rashba spin-orbit interaction, which is modulated by the geometric curvature of the 
electronic channel
to realize the topological pumping protocol originally introduced by Thouless in an entirely novel fashion. 
We expect the precise pumping of electric charges in our mesoscopic quantum device to be relevant for quantum metrology purposes. 
\end{abstract}
\date{\today}
\maketitle

\noindent 
\paragraph{ Introduction --} Precisely as in an Archimedean screw, where water is pumped by a rotating spiral tube, in a charge pump periodic perturbations induce a dc current~\cite{alt99} without an external bias. 
This phenomenon can be entirely adiabatic, as it occurs for instance in open quantum dots subject to a cyclic deformation of the confining potential~\cite{bro98,swi99}, 
with the system that always remains in its instantaneous ground state. 
In a topological charge pump~\cite{tho83} the zero bias dc current is precisely quantized, and the quantization is topologically protected against external perturbations~\cite{niu84}. 
Each pump cycle transports an integer number of electronic charges with the integer uniquely determined by a topological invariant: the Chern number $\mathcal{C}$ of the quantum system~\cite{tho82}. 
The dc current generated by periodic variations of one parameter of the system Hamiltonian amounts indeed to $I=e \, {\mathcal C} \, \nu$ where $\nu$ is the frequency of the variation~\cite{xia10}.
Topological charge pumping
can be 
also
understood as a dynamical analog of the integer quantum Hall effect~\cite{wan13,mar15}, 
with the charge pumped in each cycle that can be mapped exactly to the quantized Hall conductance of a ``dual" two-dimensional electronic system. 
Such analogy is mathematically transparent considering the mapping  
between the Hofstadter model~\cite{hof76} -- perhaps the most simple Hamiltonian to study quantum Hall physics in a two-dimensional lattice system -- and the 
one-dimensional Aubry-Andr\'e-Harper (AAH) model \cite{har55,aub80}, when the superlattice potential is assumed to rigidly slide in time \cite{wan13,mar15,lau15}. 

\begin{figure}[bp]
\includegraphics[width=.85\columnwidth]{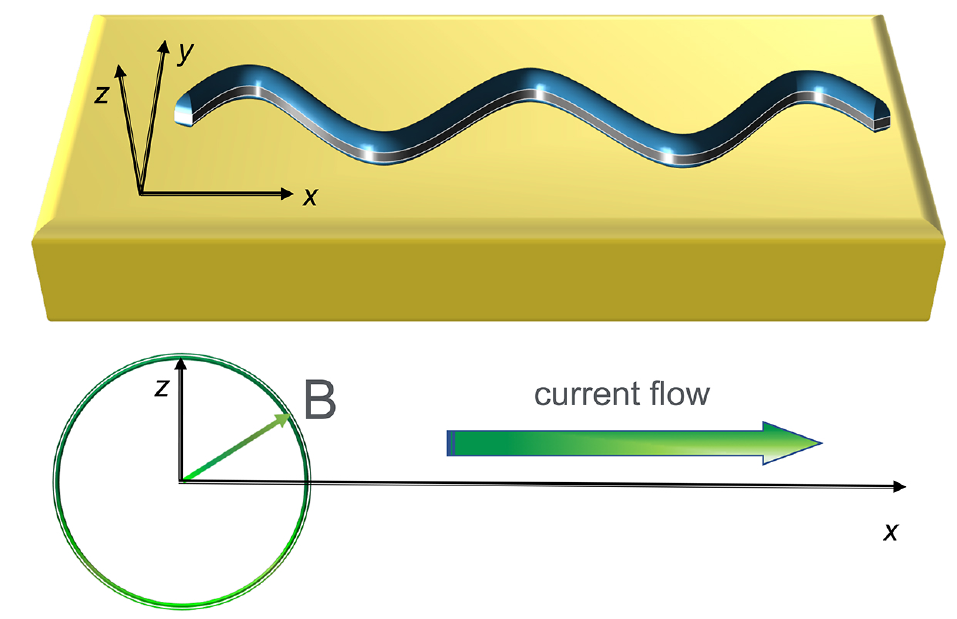}
\caption{(color online). Top: Schematic view of a semiconducting one-dimensional channel patterned in a serpentine-shape at the mesoscopic scale. 
The locally varying Rashba spin-orbit field axis lies in the $\hat{x}-\hat{y}$ plane. Bottom: Schematic diagram of the pumping induced by the rotation of the magnetic field. 
The green circle with arrow shows the magnetic field trajectory. }
\label{fig:fig1}
\end{figure}

Very recently, the advances in constructing optical lattice structures with precise control over lattice intensity and phases have enabled the realization of dynamically controlled optical superlattices, and consequently of topological quantum pumps in 
ultracold atomic systems \cite{loh16,nak16,cit16}. 
In one-dimensional electronic systems, instead, the creation of a dynamical superlattice  potential critically relies on the presence and control of superimposed oscillating local voltages~\cite{niu90}. This, in turns, severely hampers the possibility to bring topological charge pumping within reach in condensed matter experiments. 

In this Rapid Communication,  we propose and validate theoretically an entirely novel solid-state system in which  topological quantum pumping can be achieved even in the complete absence of superimposed voltage leads. The system consists of a Rashba spin-orbit coupled 
semiconducting narrow channel with a
serpentine shape at the mesoscopic scale [c.f. Fig.\ref{fig:fig1}(a)]
: It can be obtained either by processing a semiconducting quantum well lithographically~\cite{kun09}, or creating a ``zigzag" nanowire network of  crystalline quality~\cite{gaz17}.
To operate, the device makes use of an auxiliary external planar rotating magnetic field, which serves as the periodic (ac) perturbation driving the charge pumping [c.f. Fig.\ref{fig:fig1}(b)] . The concomitant presence of the time-dependent Zeeman interaction and the spin-orbit coupling, which is effectively inhomogeneous due to the geometric curvature of the 
nanostructure~\cite{gen15}
renders a sliding superlattice potential acting on the electronic charges, and ultimately yields a quantized dc current. 
We also show that as the strength of the rotating magnetic field is increased, the system undergoes a topological phase transition from a state where an even integer number of electronic charges are transported in each rotation period of the magnetic field, to a state where the rotating magnetic field does not pump any electronic charge.  

\begin{figure}[tbp]
\vspace{0 truecm}
\includegraphics[width=.75\columnwidth]{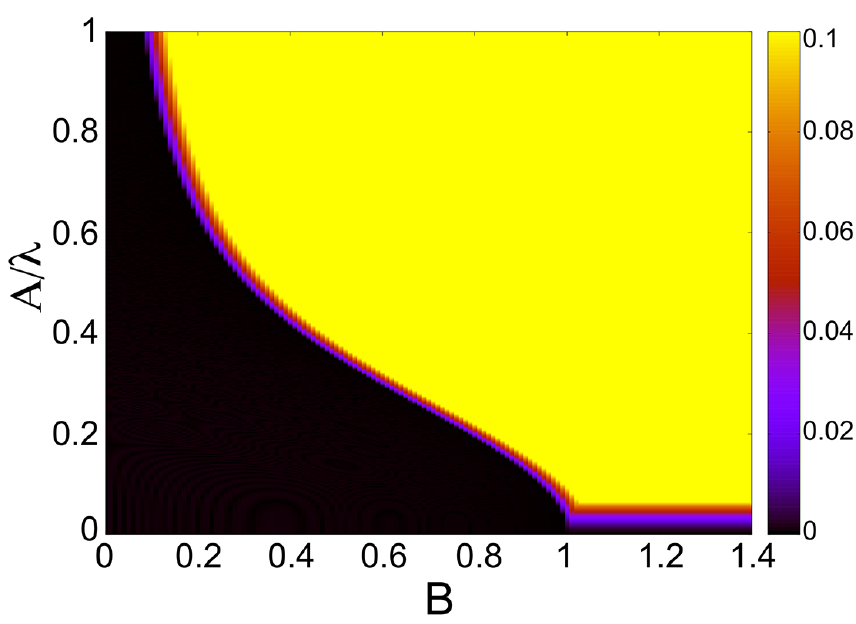}
\caption{(color online) Map of the minigap at one-quarter filling for a spin-orbit free curved channel with $\lambda = 4 a$ in the magnetic field - geometric curvature parameter space. The black region manifests the existence of a semimetallic state. The magnetic field has been measured in units of $t$.}
\label{fig:fig2}
\end{figure}

\paragraph{Theoretical model --} We start out by expressing the effective ${\bf k \cdot p}$ Hamiltonian for a planar one-dimensional 
semiconducting channel with curved geometry
as~\cite{ort15} 
\begin{eqnarray}
{\cal H}_{\bf k \cdot p}&=& - \dfrac{\hbar^2}{2 m^{\star}} \partial^2_{s} - i \alpha_{R} \tau_{{\mathcal N}} (s) \partial_s + \dfrac{i \alpha_{R}}{2} \kappa(s) \tau_{\mathcal T}(s)  \nonumber \\ & & - \dfrac{\hbar^2}{8 m^{\star}} \kappa(s)^2. 
\label{eq:hamiltonianorigin}
\end{eqnarray}
In the equation above, $s$ is the arclength along the 
channel
measured from an arbitrary reference point,  $m^{\star}$ and $\alpha_{R}$  indicate the effective mass of the carriers 
and the Rashba spin-orbit coupling constant respectively, and $\kappa(s)$ is the local curvature explicitly entering in the so-called quantum geometric potential (QGP)~\cite{jen71,dac82,ort10,ort11b}. 
We also introduced two local Pauli matrices, comoving with the electrons as they propagate along $s$, explicitly reading 
$\tau_{\mathcal N}(s) = {\boldsymbol \tau} \cdot \hat{\mathcal N}(s)$ and $\tau_{\mathcal T}(s) = {\boldsymbol \tau} \cdot \hat{\mathcal T}(s)$,
where the $\boldsymbol{\tau}$'s are the usual Pauli matrices. 
$\hat{\mathcal T}(s)$ and $\hat{\mathcal N}(s)$, {\it i.e.} the local tangential and normal directions of the narrow 
channel,
can be expressed in Euclidean space by introducing an angle $\varphi(s)$, in terms of which $\hat{\mathcal N}(s) = \left\{\sin{\varphi(s)}, \cos{\varphi(s)}, 0 \right\}$
and $\hat{\mathcal T}(s) = \left\{\cos{\varphi(s)}, -\sin{\varphi(s)}, 0 \right\}$. 
Using the Frenet-Serret  equation $\partial_s \hat{\mathcal T}(s) \equiv \kappa(s) \hat{\mathcal N}(s)$, it then follows that the angle $\varphi(s)= - \int^s \kappa(s^{\prime}) d s^{\prime}$ is entirely  determined 
by the local curvature.

To proceed further, we use that the undulating geometry of a serpentine-shape can be parametrized in the Monge gauge as $y(x) = A \cos{(2 \pi x / \lambda)}$ with $A$ 
the maximum departure of the channel from flatness, and $\lambda$ the serpentine period. 
In the shallow deformation limit $ A / \lambda \ll 1$, we can express the arclength of the channel $s \simeq x$ while the geometric curvature is 
$\kappa(s) \simeq - (\frac{2 \pi}{\lambda})^2 A \cos{(2 \pi s / \lambda)} $. This periodicity of the curvature transmutes the QGP in an effective superlattice potential~\cite{pan16}, 
and yields at the same time a periodic canting of the spin-orbit field axis~\cite{gen15}.
In order to study their concomitant effect on the electronic properties of the quantum 
system,
we next introduce a tight-binding model obtained by discretizing Eq.~\ref{eq:hamiltonianorigin} on a lattice. It can be written as 
\begin{eqnarray}
{\cal H}&=& -t  \sum_{i,\sigma} {c^{\dagger}_{i,\sigma} c_{i+1,\sigma}}  - \frac{\hbar^2}{8 m^*} \sum_{i,\sigma}{ \kappa(s_i)^2 c^{\dagger}_{i,\sigma}c_{i,\sigma}} \label{eq:hamiltonianTB}\\ & -& 
  \frac{i \alpha_R}{2} \sum_{i,\sigma,\sigma^\prime} c^{\dagger}_{i,\sigma} \left[ \frac{\tau_N^{\sigma, \sigma^{\prime}}(s_i) + \tau_N^{\sigma, \sigma^{\prime}}(s_{i+1}) }{2} \right]c_{i+1,\sigma^\prime} + H.c., \nonumber
 \end{eqnarray}
where $c^{\dagger}_{i , \sigma}$, $c_{i , \sigma}$ are operators creating and annihilating, respectively, an electron at the $i$th site with spin projection $\sigma= \uparrow, \downarrow$ along the $\hat{z}$ axis, and $t = \hbar^2 / (2 m^{\star} a^2)$ is the nearest-neighbour hopping amplitude. 
The atomic positions can be instead written as $s_i / \lambda = p\,i / q + \phi$, with $p$ and $q$ coprimes whose ratio is $p/q = a / \lambda$ ($a$ being the tight-binding lattice constant), and $\phi$ a phase accounting for nonequivalent displacements of the atoms in one superstructure period. 
The Hamiltonian in Eq.~\ref{eq:hamiltonianTB} can be seen as the combination of a conventional, commensurate diagonal AAH model with a spin-dependent off-diagonal AAH model~\cite{gan13}. Taken separately, both these models realize butterfly spectra \cite{hof76,gen15} and are characterized by insulating states with non-trivial Chern numbers if the phase $\phi$ is assumed to vary in time. 
We will instead consider a constant $\phi$ value (taking $\phi \equiv 0$ for simplicity) and monitor the effect of an external magnetic field, which we account for adding the usual Zeeman term ${\cal H}_Z \equiv \sum_{i,\sigma,\sigma^\prime}{c^{\dagger}_{i,\sigma} \left[\vec{B}\cdot\vec{\tau}\right]_{\sigma,\sigma^\prime}c_{i,\sigma^\prime}}$ to the Hamiltonian in Eq.~\ref{eq:hamiltonianTB}.

We first consider the ensuing electronic properties in the regime of negligible Rashba spin-orbit coupling, with preserved ${\cal SU}(2)$ spin symmetry, 
for the simple case of $p/q=1/4$.  At filling fraction $\nu=1/4$, the geometric curvature of the 
nanostructure
yields, via the QGP, a metal-insulator transition (MIT) for strong enough external magnetic fields. 
In this regime indeed, the lowest energy bands are completely spin-polarized thereby allowing the spin-independent superlattice potential to open up a gap at the centre of the mini-Brillouin zone (mBZ)~\cite{SuppM}. As explicitly proved in Fig.~\ref{fig:fig2}, for $A / \lambda \rightarrow 0$ the critical value of the magnetic field strength at which the MIT takes place 
is $B_c \simeq t$, and it continuously flows to smaller values as the geometric curvature is increased. In the $B<B_c$ region instead, two degeneracies between opposite spin states at unpinned points in the mBZ  cannot be removed by the action of the QGP thereby implying the existence of a semimetallic state at one-quarter filling. 

\begin{figure}[tbp]
\vspace{0 truecm}
\includegraphics[width=.75\columnwidth]{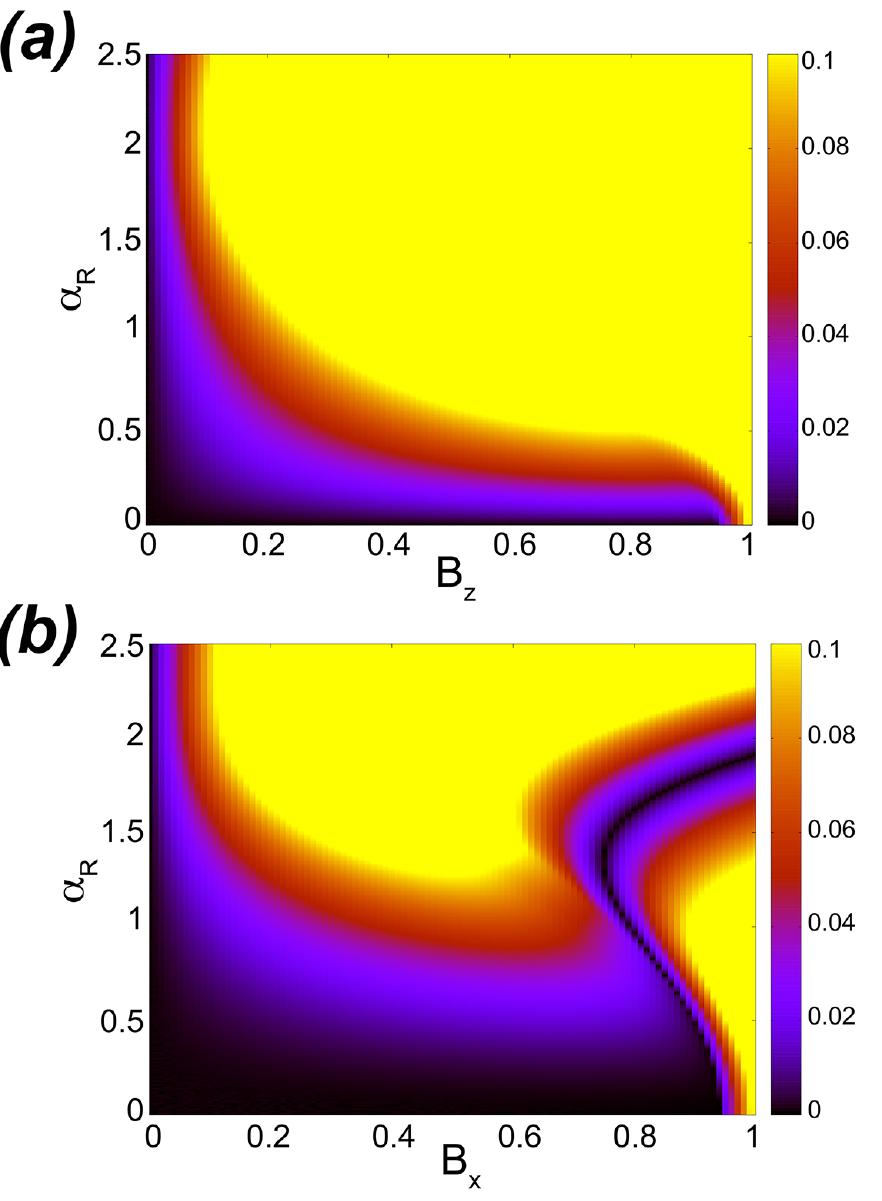}
\caption{(color online) (a). Map of the one-quarter filling gap for a serpentine-shape semiconducting channel with $A/\lambda=0.1$ as obtained by varying the Rashba strength $\alpha_R$  
and the magnetic field strength (both measured in units of $t$) assuming the direction of the magnetic field to be perpendicular to the wire plane. (b) Same for a magnetic field  oriented in the $\hat{x}$-direction [c.f. Fig.~\ref{fig:fig1}]. The black line corresponds to a closing-reopening of the gap.}
\label{fig:fig3}
\end{figure}

Having established the occurrence of a curvature-driven MIT in the presence of ${\cal SU}(2)$ spin symmetry, we now investigate the effect of the Rashba spin-orbit coupling. 
We first consider a magnetic field ${\bf B}= B ~ \hat{z}$, orthogonal to the plane in which the spin-orbit field lies. 
Contrary to a conventional 
straight channel,
the canting of the Rashba field resulting from the geometric curvature 
yields a finite gap 
at $\nu=1/4$ filling {\it independent} of the strength of the external magnetic field [c.f. Fig.~\ref{fig:fig3}(a)].  
This is because the degeneracies between opposite spin states encountered at unpinned momenta of the mBZ in the weak magnetic field regime are removed by the action of the curvature-induced modulated Rashba term. 
And indeed the gap increases monotonously with the Rashba coupling $\alpha_R$ for weak magnetic field strengths [c.f. Fig.~\ref{fig:fig3}(a)]. 
Exactly the same features  occur by tilting the magnetic field direction toward the $\hat x$ direction~\cite{SuppM}, 
which corresponds to the direction in which the effective spin-orbit interaction averages to zero in one superlattice period. 
When the magnetic field points exactly in the latter direction, however, the situation changes drastically. As shown in Fig.~\ref{fig:fig3}(b), 
cranking up the magnetic field strength leads first to an increase of the quarter filling gap, which is subsequently followed by a substantial decrease 
until a critical value where the gap undergoes a closing-reopening transition at the center of the mBZ~\cite{SuppM}.

\begin{figure}[tbp]
\includegraphics[width=\columnwidth]{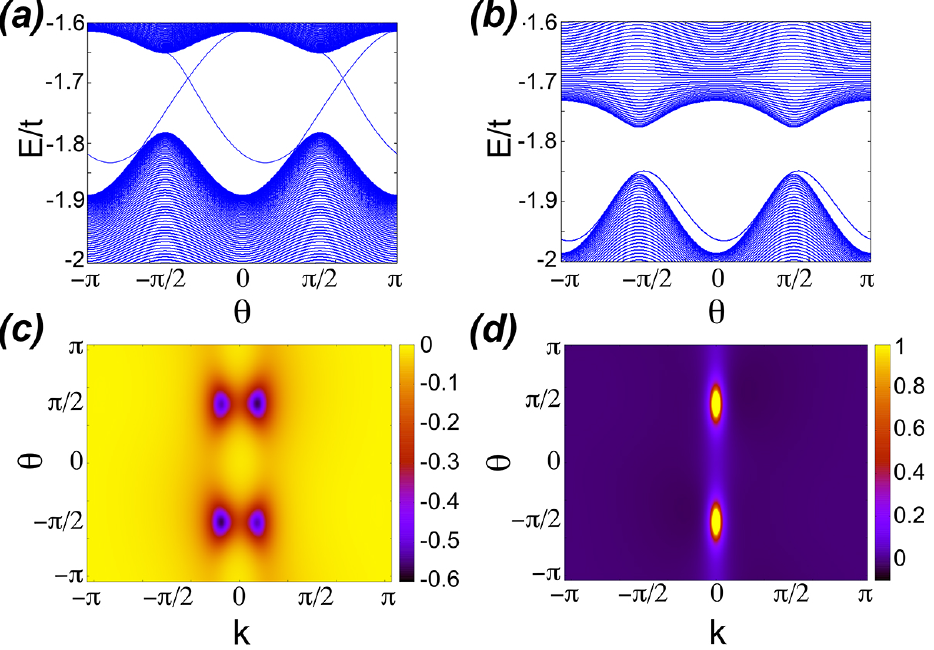}
\caption{(color online) (a)  Energy spectrum close to one-quarter filling of a serpentine-shaped semiconducting channel with a Rashba strength $\alpha_R=1.5 t$, and $A/\lambda=0.1$ subject to a magnetic field of strength $B=0.5 t$ rotating in the $\hat{x} - \hat{z}$ plane. The spectrum is obtained considering a finite atomic chain with open boundary conditions. The presence of two chiral edge modes per edge pinpoints the topological non-trivial properties of the system. (b) Same for a stronger magnetic field of strength $B= t$ where instead the system realises a conventional topologically-trivial insulator. (c),(d) The corresponding maps of the Berry curvature support the topological properties found in the weak and strong magnetic field regime respectively. In (d) the darkest regions have negative Berry curvature.}
\label{fig:fig4}
\end{figure}

\paragraph{Topological charge pumping --} In the same spirit of Thouless seminal work~\cite{tho83}, we now consider a slow, adiabatic ac rotation of the magnetic field in the $\hat{x} - \hat{z}$ plane. Since the system always remains 
in its instantaneous ground state, we can interpret the angle $\theta$ that the magnetic field direction forms with the $\hat{z}$ axis, as an additional quasi-momentum. 
With this dimensional extension, our system can be viewed as a two-dimensional insulator which, due to the absence of time-reversal symmetry, belongs to the class A of the Altland-Zirnbauer table~\cite{alt97}. 
Moreover, the presence of the gap closing-reopening discussed above suggests the existence of two insulating phases characterized by  different $\mathbb Z$ topological invariants. 
We have verified that our quarter-filled model supports an insulating phase with a non-trivial Chern number ${\mathcal C} \equiv -2$ in the weak magnetic field region, 
while a completely trivial ${\mathcal C} \equiv 0$ insulating phase is encountered in the strong-field regime. 
As shown in Fig.\ref{fig:fig4}(a), the energy spectrum of a finite size system with open boundary conditions displays  two chiral edge states 
within the one quarter-filling gap for small enough magnetic field. 
They are related to each other by a $\pi$ rotation of the magnetic field direction, and carry opposite spin content as can be shown with a symmetry argument~\cite{SuppM}.

In the opposite strong field regime, instead, the edge states do not connect the valence to the conduction band thereby implying a topologically trivial insulating state  [c.f. Fig.~\ref{fig:fig4}(b)]. 
We have additionally computed the Berry curvature of the insulating states using the method outlined in Ref.~\onlinecite{tak05}. In the weak magnetic field regime, we find that the Berry curvature displays four peaks [c.f. Fig.~\ref{fig:fig4}(c)], each of which contributing $\simeq -1/2$ to the total Chern number. In the strong magnetic field regime, instead, the contributions coming from the  two peaks located at the mBZ center for $\theta=\pm \pi/2$  [c.f. Fig.~\ref{fig:fig4}(d)], are identically cancelled by an almost homogeneous background with Berry curvature opposite in sign.

\begin{figure}
\vspace{0 truecm}
\includegraphics[width=\columnwidth]{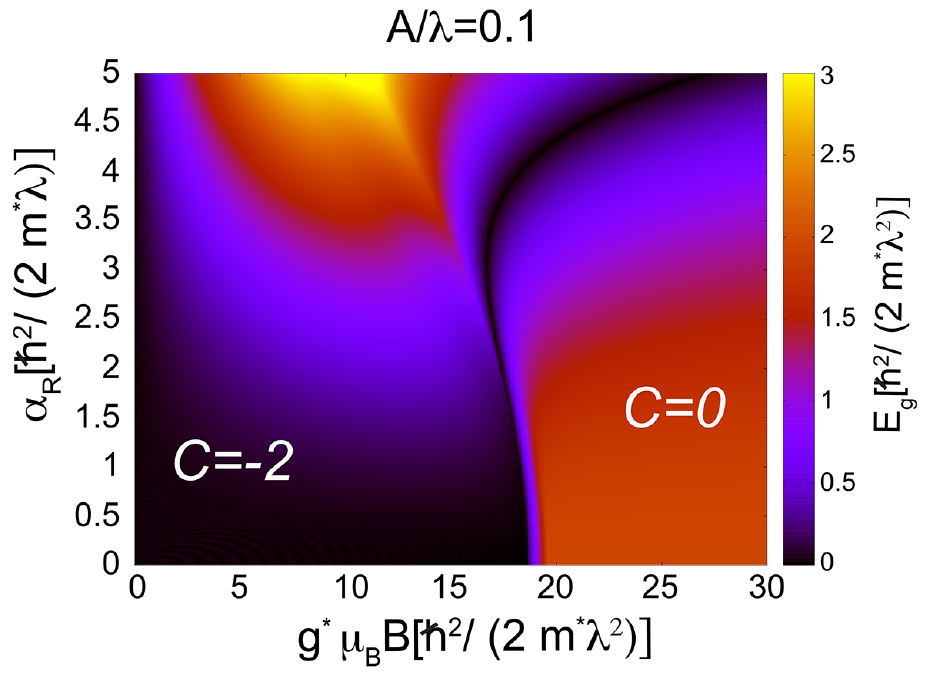}
\caption{Map of the gap between the second and third minibands of an undulating one-dimensional channel subject to rotating magnetic fields, as obtained from the continuum ${\bf k \cdot p}$ theory. The central black lines correspond to the topological phase transition separating the topologically non-trivial region (with $\mathcal{C}=-2$) at weak magnetic fields from the topologically-trivial one at higher magnetic fields. Energies are measured in units of $\hbar^2 / (2 m^{\star} \lambda^2)$.}
\label{fig:fig5}
\end{figure}

The fact that our tight-binding model realises a $\mathcal{C} \equiv -2$ topological charge pump when subject to weak rotating magnetic fields is independent of the superlattice period. Specifically, 
for any superlattice period $\lambda= 2 n a$ with $n$ integer, and at $1 / ( 2 n)$ filling, all the discussion above remains unaltered. This, in turn, implies that also a  mesoscale serpentine-shaped quantum wire 
can operate as a topological charge pump. 
Since the conditions for this effect to be observed are $k_B ~ T$, $h \nu < E_{g}$ where $k_B ~ T$  is the energy scale for temperature $T$, 
$\nu$ the magnetic field frequency,
and $E_g$ is the relevant energy gap, we analyzed the continuum ${\bf k \cdot p}$ effective theory given by Eq.~\ref{eq:hamiltonianorigin} taking into account 
the Zeeman term ${\cal H}_Z= g^{\star} \mu_B {\bf B} \cdot {\boldsymbol \tau}$, where $g^{\star}$ is the effective Land\'e $g$ factor, 
and $\mu_B$ the Bohr magneton. 
Fig.~\ref{fig:fig5} shows the corresponding topological phase diagrams with a map of the bulk gap when the first two Bloch minibands are fully occupied. 
By considering the serpertine period to be of the order of 
$\lambda \simeq 0.5~\mu$m,
the effective mass of, {\it e.g.}, InAs $m^{\star} \simeq 0.023$, and a strength of the Rashba spin-orbit interaction~\cite{lia12} up to 
$0.2$~eV~\AA,
we find that the size of the bulk gap 
$E_{g} \simeq 20~\mu$eV
for $A/\lambda=0.1$ and a magnetic field strength 
$\simeq 60$mT
(the Land\'e $g$ factor $g_{InAs} \simeq 15$ in bulk crystals~\cite{ali12}). 
As a result, a topological charge pumping effect characterised by a quantized dc current can be observed for temperatures 
$T < 200~$mK
and frequencies 
$\omega < 4$~GHz. A rotating magnetic field with this freqency can be in principle obtained by running current pulses in two perpendicular conductors with a $\pi/2$ phase shift~\cite{li16}. The resulting dc current $I \simeq 1~$nA can be easily detected with present day experimental capabilities. Therefore,
in this physical regime our solid-state device can be used to measure the fundamental electron charge $e$.

\paragraph{Conclusions --} To wrap up, we have shown that a Rashba spin-orbit coupled 
semiconducting channel possessing a mesoscale serpentine shape, as obtained either by lithographic processing~\cite{kun09} or using ``zigzag" nanowire networks~\cite{gaz17},
can operate as a topological charge pump once subject to a relatively weak magnetic field rotating on a plane. 
Being topological in nature, the electronic pumping is robust against additional perturbations, {\it e.g.} tiny tilts of the magnetic field in the direction orthogonal to the plane. 
The existence of this phenomenon stems from the 
periodic canting of the Rashba spin-orbit field, which is the prime physical consequence of the geometric curvature of our 
nanostructure,
and provides a bridge between the Zeeman 
interaction and the quantized charge flow in the wire. 
Adiabatic quantum pumping has been recently discussed in one dimensional semiconducting Rashba systems in the presence of local voltage leads or periodically arranged nanomagnets~\cite{sah14,rai14}. Our proposed set-up, instead, does not necessitate the usage of external local perturbation 
since it only relies on the successful 
growth of a one-dimensional channel with curved undulating geometry. 
Our findings therefore add a prime actor to the recently uncovered series of unique curvature-induced quantum effects in low-dimensional semiconducting systems~\cite{cha14,yin16,cha17}. 

\paragraph{Acknowledgements --} We acknowledge the financial support of the Future and Emerging Technologies (FET) programme under FET-Open grant number: 618083 (CNTQC). 
C.O. acknowledges support from the Deutsche Forschungsgemeinschaft (Grant No. OR 404/1-1), and from a VIDI grant (Project 680-47-543) financed by the Netherlands Organization for Scientific Research (NWO).
S.~P. thanks Ulrike Nitzsche for technical assistance.
N.~S. thanks Guido van Miert for many fruitful discussions.
S.~P. and N.~S. contributed equally to this work.


\begin{thebibliography}{37}%
\makeatletter
\providecommand \@ifxundefined [1]{%
 \@ifx{#1\undefined}
}%
\providecommand \@ifnum [1]{%
 \ifnum #1\expandafter \@firstoftwo
 \else \expandafter \@secondoftwo
 \fi
}%
\providecommand \@ifx [1]{%
 \ifx #1\expandafter \@firstoftwo
 \else \expandafter \@secondoftwo
 \fi
}%
\providecommand \natexlab [1]{#1}%
\providecommand \enquote  [1]{``#1''}%
\providecommand \bibnamefont  [1]{#1}%
\providecommand \bibfnamefont [1]{#1}%
\providecommand \citenamefont [1]{#1}%
\providecommand \href@noop [0]{\@secondoftwo}%
\providecommand \href [0]{\begingroup \@sanitize@url \@href}%
\providecommand \@href[1]{\@@startlink{#1}\@@href}%
\providecommand \@@href[1]{\endgroup#1\@@endlink}%
\providecommand \@sanitize@url [0]{\catcode `\\12\catcode `\$12\catcode
  `\&12\catcode `\#12\catcode `\^12\catcode `\_12\catcode `\%12\relax}%
\providecommand \@@startlink[1]{}%
\providecommand \@@endlink[0]{}%
\providecommand \url  [0]{\begingroup\@sanitize@url \@url }%
\providecommand \@url [1]{\endgroup\@href {#1}{\urlprefix }}%
\providecommand \urlprefix  [0]{URL }%
\providecommand \Eprint [0]{\href }%
\providecommand \doibase [0]{http://dx.doi.org/}%
\providecommand \selectlanguage [0]{\@gobble}%
\providecommand \bibinfo  [0]{\@secondoftwo}%
\providecommand \bibfield  [0]{\@secondoftwo}%
\providecommand \translation [1]{[#1]}%
\providecommand \BibitemOpen [0]{}%
\providecommand \bibitemStop [0]{}%
\providecommand \bibitemNoStop [0]{.\EOS\space}%
\providecommand \EOS [0]{\spacefactor3000\relax}%
\providecommand \BibitemShut  [1]{\csname bibitem#1\endcsname}%
\let\auto@bib@innerbib\@empty
\bibitem [{\citenamefont {Altshuler}\ and\ \citenamefont
  {Glazman}(1999)}]{alt99}%
  \BibitemOpen
  \bibfield  {author} {\bibinfo {author} {\bibfnamefont {B.~L.}\ \bibnamefont
  {Altshuler}}\ and\ \bibinfo {author} {\bibfnamefont {L.~I.}\ \bibnamefont
  {Glazman}},\ }\href {\doibase 10.1126/science.283.5409.1864} {\bibfield
  {journal} {\bibinfo  {journal} {Science}\ }\textbf {\bibinfo {volume}
  {283}},\ \bibinfo {pages} {1864} (\bibinfo {year} {1999})}\BibitemShut
  {NoStop}%
\bibitem [{\citenamefont {Brouwer}(1998)}]{bro98}%
  \BibitemOpen
  \bibfield  {author} {\bibinfo {author} {\bibfnamefont {P.~W.}\ \bibnamefont
  {Brouwer}},\ }\href {\doibase 10.1103/PhysRevB.58.R10135} {\bibfield
  {journal} {\bibinfo  {journal} {Phys. Rev. B}\ }\textbf {\bibinfo {volume}
  {58}},\ \bibinfo {pages} {R10135} (\bibinfo {year} {1998})}\BibitemShut
  {NoStop}%
\bibitem [{\citenamefont {Switkes}\ \emph {et~al.}(1999)\citenamefont
  {Switkes}, \citenamefont {Marcus}, \citenamefont {Campman},\ and\
  \citenamefont {Gossard}}]{swi99}%
  \BibitemOpen
  \bibfield  {author} {\bibinfo {author} {\bibfnamefont {M.}~\bibnamefont
  {Switkes}}, \bibinfo {author} {\bibfnamefont {C.~M.}\ \bibnamefont {Marcus}},
  \bibinfo {author} {\bibfnamefont {K.}~\bibnamefont {Campman}}, \ and\
  \bibinfo {author} {\bibfnamefont {A.~C.}\ \bibnamefont {Gossard}},\ }\href
  {\doibase 10.1126/science.283.5409.1905} {\bibfield  {journal} {\bibinfo
  {journal} {Science}\ }\textbf {\bibinfo {volume} {283}},\ \bibinfo {pages}
  {1905} (\bibinfo {year} {1999})}\BibitemShut {NoStop}%
\bibitem [{\citenamefont {Thouless}(1983)}]{tho83}%
  \BibitemOpen
  \bibfield  {author} {\bibinfo {author} {\bibfnamefont {D.~J.}\ \bibnamefont
  {Thouless}},\ }\href {\doibase 10.1103/PhysRevB.27.6083} {\bibfield
  {journal} {\bibinfo  {journal} {Phys. Rev. B}\ }\textbf {\bibinfo {volume}
  {27}},\ \bibinfo {pages} {6083} (\bibinfo {year} {1983})}\BibitemShut
  {NoStop}%
\bibitem [{\citenamefont {Niu}\ and\ \citenamefont {Thouless}(1984)}]{niu84}%
  \BibitemOpen
  \bibfield  {author} {\bibinfo {author} {\bibfnamefont {Q.}~\bibnamefont
  {Niu}}\ and\ \bibinfo {author} {\bibfnamefont {D.~J.}\ \bibnamefont
  {Thouless}},\ }\href@noop {} {\bibfield  {journal} {\bibinfo  {journal}
  {Journal of Physics A: Mathematical and General}\ }\textbf {\bibinfo {volume}
  {17}},\ \bibinfo {pages} {2453} (\bibinfo {year} {1984})}\BibitemShut
  {NoStop}%
\bibitem [{\citenamefont {Thouless}\ \emph {et~al.}(1982)\citenamefont
  {Thouless}, \citenamefont {Kohmoto}, \citenamefont {Nightingale},\ and\
  \citenamefont {den Nijs}}]{tho82}%
  \BibitemOpen
  \bibfield  {author} {\bibinfo {author} {\bibfnamefont {D.~J.}\ \bibnamefont
  {Thouless}}, \bibinfo {author} {\bibfnamefont {M.}~\bibnamefont {Kohmoto}},
  \bibinfo {author} {\bibfnamefont {M.~P.}\ \bibnamefont {Nightingale}}, \ and\
  \bibinfo {author} {\bibfnamefont {M.}~\bibnamefont {den Nijs}},\ }\href
  {\doibase 10.1103/PhysRevLett.49.405} {\bibfield  {journal} {\bibinfo
  {journal} {Phys. Rev. Lett.}\ }\textbf {\bibinfo {volume} {49}},\ \bibinfo
  {pages} {405} (\bibinfo {year} {1982})}\BibitemShut {NoStop}%
\bibitem [{\citenamefont {Xiao}\ \emph {et~al.}(2010)\citenamefont {Xiao},
  \citenamefont {Chang},\ and\ \citenamefont {Niu}}]{xia10}%
  \BibitemOpen
  \bibfield  {author} {\bibinfo {author} {\bibfnamefont {D.}~\bibnamefont
  {Xiao}}, \bibinfo {author} {\bibfnamefont {M.-C.}\ \bibnamefont {Chang}}, \
  and\ \bibinfo {author} {\bibfnamefont {Q.}~\bibnamefont {Niu}},\ }\href
  {\doibase 10.1103/RevModPhys.82.1959} {\bibfield  {journal} {\bibinfo
  {journal} {Rev. Mod. Phys.}\ }\textbf {\bibinfo {volume} {82}},\ \bibinfo
  {pages} {1959} (\bibinfo {year} {2010})}\BibitemShut {NoStop}%
\bibitem [{\citenamefont {Wang}\ \emph {et~al.}(2013)\citenamefont {Wang},
  \citenamefont {Troyer},\ and\ \citenamefont {Dai}}]{wan13}%
  \BibitemOpen
  \bibfield  {author} {\bibinfo {author} {\bibfnamefont {L.}~\bibnamefont
  {Wang}}, \bibinfo {author} {\bibfnamefont {M.}~\bibnamefont {Troyer}}, \ and\
  \bibinfo {author} {\bibfnamefont {X.}~\bibnamefont {Dai}},\ }\href {\doibase
  10.1103/PhysRevLett.111.026802} {\bibfield  {journal} {\bibinfo  {journal}
  {Phys. Rev. Lett.}\ }\textbf {\bibinfo {volume} {111}},\ \bibinfo {pages}
  {026802} (\bibinfo {year} {2013})}\BibitemShut {NoStop}%
\bibitem [{\citenamefont {Marra}\ \emph {et~al.}(2015)\citenamefont {Marra},
  \citenamefont {Citro},\ and\ \citenamefont {Ortix}}]{mar15}%
  \BibitemOpen
  \bibfield  {author} {\bibinfo {author} {\bibfnamefont {P.}~\bibnamefont
  {Marra}}, \bibinfo {author} {\bibfnamefont {R.}~\bibnamefont {Citro}}, \ and\
  \bibinfo {author} {\bibfnamefont {C.}~\bibnamefont {Ortix}},\ }\href
  {\doibase 10.1103/PhysRevB.91.125411} {\bibfield  {journal} {\bibinfo
  {journal} {Phys. Rev. B}\ }\textbf {\bibinfo {volume} {91}},\ \bibinfo
  {pages} {125411} (\bibinfo {year} {2015})}\BibitemShut {NoStop}%
\bibitem [{\citenamefont {Hofstadter}(1976)}]{hof76}%
  \BibitemOpen
  \bibfield  {author} {\bibinfo {author} {\bibfnamefont {D.~R.}\ \bibnamefont
  {Hofstadter}},\ }\href {\doibase 10.1103/PhysRevB.14.2239} {\bibfield
  {journal} {\bibinfo  {journal} {Phys. Rev. B}\ }\textbf {\bibinfo {volume}
  {14}},\ \bibinfo {pages} {2239} (\bibinfo {year} {1976})}\BibitemShut
  {NoStop}%
\bibitem [{\citenamefont {Harper}(1955)}]{har55}%
  \BibitemOpen
  \bibfield  {author} {\bibinfo {author} {\bibfnamefont {P.~G.}\ \bibnamefont
  {Harper}},\ }\href@noop {} {\bibfield  {journal} {\bibinfo  {journal}
  {Proceedings of the Physical Society. Section A}\ }\textbf {\bibinfo {volume}
  {68}},\ \bibinfo {pages} {874} (\bibinfo {year} {1955})}\BibitemShut
  {NoStop}%
\bibitem [{\citenamefont {Aubry}\ and\ \citenamefont {Andr\'e}(1980)}]{aub80}%
  \BibitemOpen
  \bibfield  {author} {\bibinfo {author} {\bibfnamefont {S.}~\bibnamefont
  {Aubry}}\ and\ \bibinfo {author} {\bibfnamefont {G.}~\bibnamefont
  {Andr\'e}},\ }\href@noop {} {\bibfield  {journal} {\bibinfo  {journal} {Ann.
  Isr. Phys. Soc.}\ }\textbf {\bibinfo {volume} {3}},\ \bibinfo {pages} {133}
  (\bibinfo {year} {1980})}\BibitemShut {NoStop}%
\bibitem [{\citenamefont {Lau}\ \emph {et~al.}(2015)\citenamefont {Lau},
  \citenamefont {Ortix},\ and\ \citenamefont {van~den Brink}}]{lau15}%
  \BibitemOpen
  \bibfield  {author} {\bibinfo {author} {\bibfnamefont {A.}~\bibnamefont
  {Lau}}, \bibinfo {author} {\bibfnamefont {C.}~\bibnamefont {Ortix}}, \ and\
  \bibinfo {author} {\bibfnamefont {J.}~\bibnamefont {van~den Brink}},\ }\href
  {\doibase 10.1103/PhysRevLett.115.216805} {\bibfield  {journal} {\bibinfo
  {journal} {Phys. Rev. Lett.}\ }\textbf {\bibinfo {volume} {115}},\ \bibinfo
  {pages} {216805} (\bibinfo {year} {2015})}\BibitemShut {NoStop}%
\bibitem [{\citenamefont {Lohse}\ \emph {et~al.}(2016)\citenamefont {Lohse},
  \citenamefont {Schweizer}, \citenamefont {Zilberberg}, \citenamefont
  {Aidelsburger},\ and\ \citenamefont {Bloch}}]{loh16}%
  \BibitemOpen
  \bibfield  {author} {\bibinfo {author} {\bibfnamefont {M.}~\bibnamefont
  {Lohse}}, \bibinfo {author} {\bibfnamefont {C.}~\bibnamefont {Schweizer}},
  \bibinfo {author} {\bibfnamefont {O.}~\bibnamefont {Zilberberg}}, \bibinfo
  {author} {\bibfnamefont {M.}~\bibnamefont {Aidelsburger}}, \ and\ \bibinfo
  {author} {\bibfnamefont {I.}~\bibnamefont {Bloch}},\ }\href@noop {}
  {\bibfield  {journal} {\bibinfo  {journal} {Nat Phys}\ }\textbf {\bibinfo
  {volume} {12}},\ \bibinfo {pages} {350} (\bibinfo {year} {2016})}\BibitemShut
  {NoStop}%
\bibitem [{\citenamefont {Nakajima}\ \emph {et~al.}(2016)\citenamefont
  {Nakajima}, \citenamefont {Tomita}, \citenamefont {Taie}, \citenamefont
  {Ichinose}, \citenamefont {Ozawa}, \citenamefont {Wang}, \citenamefont
  {Troyer},\ and\ \citenamefont {Takahashi}}]{nak16}%
  \BibitemOpen
  \bibfield  {author} {\bibinfo {author} {\bibfnamefont {S.}~\bibnamefont
  {Nakajima}}, \bibinfo {author} {\bibfnamefont {T.}~\bibnamefont {Tomita}},
  \bibinfo {author} {\bibfnamefont {S.}~\bibnamefont {Taie}}, \bibinfo {author}
  {\bibfnamefont {T.}~\bibnamefont {Ichinose}}, \bibinfo {author}
  {\bibfnamefont {H.}~\bibnamefont {Ozawa}}, \bibinfo {author} {\bibfnamefont
  {L.}~\bibnamefont {Wang}}, \bibinfo {author} {\bibfnamefont {M.}~\bibnamefont
  {Troyer}}, \ and\ \bibinfo {author} {\bibfnamefont {Y.}~\bibnamefont
  {Takahashi}},\ }\href@noop {} {\bibfield  {journal} {\bibinfo  {journal} {Nat
  Phys}\ }\textbf {\bibinfo {volume} {12}},\ \bibinfo {pages} {296} (\bibinfo
  {year} {2016})}\BibitemShut {NoStop}%
\bibitem [{\citenamefont {Citro}(2016)}]{cit16}%
  \BibitemOpen
  \bibfield  {author} {\bibinfo {author} {\bibfnamefont {R.}~\bibnamefont
  {Citro}},\ }\href@noop {} {\bibfield  {journal} {\bibinfo  {journal} {Nat
  Phys}\ }\textbf {\bibinfo {volume} {12}},\ \bibinfo {pages} {288} (\bibinfo
  {year} {2016})}\BibitemShut {NoStop}%
\bibitem [{\citenamefont {Niu}(1990)}]{niu90}%
  \BibitemOpen
  \bibfield  {author} {\bibinfo {author} {\bibfnamefont {Q.}~\bibnamefont
  {Niu}},\ }\href {\doibase 10.1103/PhysRevLett.64.1812} {\bibfield  {journal}
  {\bibinfo  {journal} {Phys. Rev. Lett.}\ }\textbf {\bibinfo {volume} {64}},\
  \bibinfo {pages} {1812} (\bibinfo {year} {1990})}\BibitemShut {NoStop}%
\bibitem [{\citenamefont {Kunihashi}\ \emph {et~al.}(2009)\citenamefont
  {Kunihashi}, \citenamefont {Kohda},\ and\ \citenamefont {Nitta}}]{kun09}%
  \BibitemOpen
  \bibfield  {author} {\bibinfo {author} {\bibfnamefont {Y.}~\bibnamefont
  {Kunihashi}}, \bibinfo {author} {\bibfnamefont {M.}~\bibnamefont {Kohda}}, \
  and\ \bibinfo {author} {\bibfnamefont {J.}~\bibnamefont {Nitta}},\ }\href
  {\doibase 10.1103/PhysRevLett.102.226601} {\bibfield  {journal} {\bibinfo
  {journal} {Phys. Rev. Lett.}\ }\textbf {\bibinfo {volume} {102}},\ \bibinfo
  {pages} {226601} (\bibinfo {year} {2009})}\BibitemShut {NoStop}%
\bibitem [{\citenamefont {Gazibegovic}\ \emph {et~al.}(2017)\citenamefont
  {Gazibegovic}, \citenamefont {Car}, \citenamefont {Zhang}, \citenamefont
  {Balk}, \citenamefont {Logan}, \citenamefont {de~Moor}, \citenamefont
  {Cassidy}, \citenamefont {Schmits}, \citenamefont {Xu}, \citenamefont {Wang},
  \citenamefont {Krogstrup}, \citenamefont {Op~het Veld}, \citenamefont {Zuo},
  \citenamefont {Vos}, \citenamefont {Shen}, \citenamefont {Bouman},
  \citenamefont {Shojaei}, \citenamefont {Pennachio}, \citenamefont {Lee},
  \citenamefont {van Veldhoven}, \citenamefont {Koelling}, \citenamefont
  {Verheijen}, \citenamefont {Kouwenhoven}, \citenamefont {Palmstr{\o}m},\ and\
  \citenamefont {Bakkers}}]{gaz17}%
  \BibitemOpen
  \bibfield  {author} {\bibinfo {author} {\bibfnamefont {S.}~\bibnamefont
  {Gazibegovic}}, \bibinfo {author} {\bibfnamefont {D.}~\bibnamefont {Car}},
  \bibinfo {author} {\bibfnamefont {H.}~\bibnamefont {Zhang}}, \bibinfo
  {author} {\bibfnamefont {S.~C.}\ \bibnamefont {Balk}}, \bibinfo {author}
  {\bibfnamefont {J.~A.}\ \bibnamefont {Logan}}, \bibinfo {author}
  {\bibfnamefont {M.~W.~A.}\ \bibnamefont {de~Moor}}, \bibinfo {author}
  {\bibfnamefont {M.~C.}\ \bibnamefont {Cassidy}}, \bibinfo {author}
  {\bibfnamefont {R.}~\bibnamefont {Schmits}}, \bibinfo {author} {\bibfnamefont
  {D.}~\bibnamefont {Xu}}, \bibinfo {author} {\bibfnamefont {G.}~\bibnamefont
  {Wang}}, \bibinfo {author} {\bibfnamefont {P.}~\bibnamefont {Krogstrup}},
  \bibinfo {author} {\bibfnamefont {R.~L.~M.}\ \bibnamefont {Op~het Veld}},
  \bibinfo {author} {\bibfnamefont {K.}~\bibnamefont {Zuo}}, \bibinfo {author}
  {\bibfnamefont {Y.}~\bibnamefont {Vos}}, \bibinfo {author} {\bibfnamefont
  {J.}~\bibnamefont {Shen}}, \bibinfo {author} {\bibfnamefont {D.}~\bibnamefont
  {Bouman}}, \bibinfo {author} {\bibfnamefont {B.}~\bibnamefont {Shojaei}},
  \bibinfo {author} {\bibfnamefont {D.}~\bibnamefont {Pennachio}}, \bibinfo
  {author} {\bibfnamefont {J.~S.}\ \bibnamefont {Lee}}, \bibinfo {author}
  {\bibfnamefont {P.~J.}\ \bibnamefont {van Veldhoven}}, \bibinfo {author}
  {\bibfnamefont {S.}~\bibnamefont {Koelling}}, \bibinfo {author}
  {\bibfnamefont {M.~A.}\ \bibnamefont {Verheijen}}, \bibinfo {author}
  {\bibfnamefont {L.~P.}\ \bibnamefont {Kouwenhoven}}, \bibinfo {author}
  {\bibfnamefont {C.~J.}\ \bibnamefont {Palmstr{\o}m}}, \ and\ \bibinfo
  {author} {\bibfnamefont {E.~P. A.~M.}\ \bibnamefont {Bakkers}},\ }\href
  {http://dx.doi.org/10.1038/nature23468} {\bibfield  {journal} {\bibinfo
  {journal} {Nature}\ }\textbf {\bibinfo {volume} {548}},\ \bibinfo {pages}
  {434} (\bibinfo {year} {2017})}\BibitemShut {NoStop}%
\bibitem [{\citenamefont {Gentile}\ \emph {et~al.}(2015)\citenamefont
  {Gentile}, \citenamefont {Cuoco},\ and\ \citenamefont {Ortix}}]{gen15}%
  \BibitemOpen
  \bibfield  {author} {\bibinfo {author} {\bibfnamefont {P.}~\bibnamefont
  {Gentile}}, \bibinfo {author} {\bibfnamefont {M.}~\bibnamefont {Cuoco}}, \
  and\ \bibinfo {author} {\bibfnamefont {C.}~\bibnamefont {Ortix}},\ }\href
  {\doibase 10.1103/PhysRevLett.115.256801} {\bibfield  {journal} {\bibinfo
  {journal} {Phys. Rev. Lett.}\ }\textbf {\bibinfo {volume} {115}},\ \bibinfo
  {pages} {256801} (\bibinfo {year} {2015})}\BibitemShut {NoStop}%
\bibitem [{\citenamefont {Ortix}(2015)}]{ort15}%
  \BibitemOpen
  \bibfield  {author} {\bibinfo {author} {\bibfnamefont {C.}~\bibnamefont
  {Ortix}},\ }\href {\doibase 10.1103/PhysRevB.91.245412} {\bibfield  {journal}
  {\bibinfo  {journal} {Phys. Rev. B}\ }\textbf {\bibinfo {volume} {91}},\
  \bibinfo {pages} {245412} (\bibinfo {year} {2015})}\BibitemShut {NoStop}%
\bibitem [{\citenamefont {Jensen}\ and\ \citenamefont {Koppe}(1971)}]{jen71}%
  \BibitemOpen
  \bibfield  {author} {\bibinfo {author} {\bibfnamefont {H.}~\bibnamefont
  {Jensen}}\ and\ \bibinfo {author} {\bibfnamefont {H.}~\bibnamefont {Koppe}},\
  }\href@noop {} {\bibfield  {journal} {\bibinfo  {journal} {Ann. Phys.}\
  }\textbf {\bibinfo {volume} {63}},\ \bibinfo {pages} {586} (\bibinfo {year}
  {1971})}\BibitemShut {NoStop}%
\bibitem [{\citenamefont {da~Costa}(1981)}]{dac82}%
  \BibitemOpen
  \bibfield  {author} {\bibinfo {author} {\bibfnamefont {R.~C.~T.}\
  \bibnamefont {da~Costa}},\ }\href {\doibase 10.1103/PhysRevA.23.1982}
  {\bibfield  {journal} {\bibinfo  {journal} {Phys.\ Rev. \ A}\ }\textbf
  {\bibinfo {volume} {23}},\ \bibinfo {pages} {1982} (\bibinfo {year}
  {1981})}\BibitemShut {NoStop}%
\bibitem [{\citenamefont {Ortix}\ and\ \citenamefont {van~den
  Brink}(2010)}]{ort10}%
  \BibitemOpen
  \bibfield  {author} {\bibinfo {author} {\bibfnamefont {C.}~\bibnamefont
  {Ortix}}\ and\ \bibinfo {author} {\bibfnamefont {J.}~\bibnamefont {van~den
  Brink}},\ }\href {\doibase 10.1103/PhysRevB.81.165419} {\bibfield  {journal}
  {\bibinfo  {journal} {Phys. Rev. B}\ }\textbf {\bibinfo {volume} {81}},\
  \bibinfo {pages} {165419} (\bibinfo {year} {2010})}\BibitemShut {NoStop}%
\bibitem [{\citenamefont {Ortix}\ \emph {et~al.}(2011)\citenamefont {Ortix},
  \citenamefont {Kiravittaya}, \citenamefont {Schmidt},\ and\ \citenamefont
  {van~den Brink}}]{ort11b}%
  \BibitemOpen
  \bibfield  {author} {\bibinfo {author} {\bibfnamefont {C.}~\bibnamefont
  {Ortix}}, \bibinfo {author} {\bibfnamefont {S.}~\bibnamefont {Kiravittaya}},
  \bibinfo {author} {\bibfnamefont {O.~G.}\ \bibnamefont {Schmidt}}, \ and\
  \bibinfo {author} {\bibfnamefont {J.}~\bibnamefont {van~den Brink}},\ }\href
  {\doibase 10.1103/PhysRevB.84.045438} {\bibfield  {journal} {\bibinfo
  {journal} {Phys. Rev. B}\ }\textbf {\bibinfo {volume} {84}},\ \bibinfo
  {pages} {045438} (\bibinfo {year} {2011})}\BibitemShut {NoStop}%
\bibitem [{\citenamefont {Pandey}\ and\ \citenamefont {Ortix}(2016)}]{pan16}%
  \BibitemOpen
  \bibfield  {author} {\bibinfo {author} {\bibfnamefont {S.}~\bibnamefont
  {Pandey}}\ and\ \bibinfo {author} {\bibfnamefont {C.}~\bibnamefont {Ortix}},\
  }\href {\doibase 10.1103/PhysRevB.93.195420} {\bibfield  {journal} {\bibinfo
  {journal} {Phys. Rev. B}\ }\textbf {\bibinfo {volume} {93}},\ \bibinfo
  {pages} {195420} (\bibinfo {year} {2016})}\BibitemShut {NoStop}%
\bibitem [{\citenamefont {Ganeshan}\ \emph {et~al.}(2013)\citenamefont
  {Ganeshan}, \citenamefont {Sun},\ and\ \citenamefont {Das~Sarma}}]{gan13}%
  \BibitemOpen
  \bibfield  {author} {\bibinfo {author} {\bibfnamefont {S.}~\bibnamefont
  {Ganeshan}}, \bibinfo {author} {\bibfnamefont {K.}~\bibnamefont {Sun}}, \
  and\ \bibinfo {author} {\bibfnamefont {S.}~\bibnamefont {Das~Sarma}},\ }\href
  {\doibase 10.1103/PhysRevLett.110.180403} {\bibfield  {journal} {\bibinfo
  {journal} {Phys. Rev. Lett.}\ }\textbf {\bibinfo {volume} {110}},\ \bibinfo
  {pages} {180403} (\bibinfo {year} {2013})}\BibitemShut {NoStop}%
\bibitem{SuppM} See Supplemental Material at [..URL..] for the electronic bandstructure in the presence and absence of Rashba spin-orbit interaction, 
the minigap behavior for a different orientation of the magnetic field, and the topological phase diagram as obtained from the continuum model increasing the geometric curvature.
\bibitem [{\citenamefont {Altland}\ and\ \citenamefont
  {Zirnbauer}(1997)}]{alt97}%
  \BibitemOpen
  \bibfield  {author} {\bibinfo {author} {\bibfnamefont {A.}~\bibnamefont
  {Altland}}\ and\ \bibinfo {author} {\bibfnamefont {M.~R.}\ \bibnamefont
  {Zirnbauer}},\ }\href {\doibase 10.1103/PhysRevB.55.1142} {\bibfield
  {journal} {\bibinfo  {journal} {Phys. Rev. B}\ }\textbf {\bibinfo {volume}
  {55}},\ \bibinfo {pages} {1142} (\bibinfo {year} {1997})}\BibitemShut
  {NoStop}%
\bibitem [{\citenamefont {Fukui}\ \emph {et~al.}(2005)\citenamefont {Fukui},
  \citenamefont {Hatsugai},\ and\ \citenamefont {Suzuki}}]{tak05}%
  \BibitemOpen
  \bibfield  {author} {\bibinfo {author} {\bibfnamefont {T.}~\bibnamefont
  {Fukui}}, \bibinfo {author} {\bibfnamefont {Y.}~\bibnamefont {Hatsugai}}, \
  and\ \bibinfo {author} {\bibfnamefont {H.}~\bibnamefont {Suzuki}},\ }\href
  {\doibase 10.1143/JPSJ.74.1674} {\bibfield  {journal} {\bibinfo  {journal}
  {Journal of the Physical Society of Japan}\ }\textbf {\bibinfo {volume}
  {74}},\ \bibinfo {pages} {1674} (\bibinfo {year} {2005})}\BibitemShut
  {NoStop}%
\bibitem [{\citenamefont {Liang}\ and\ \citenamefont {Gao}(2012)}]{lia12}%
  \BibitemOpen
  \bibfield  {author} {\bibinfo {author} {\bibfnamefont {D.}~\bibnamefont
  {Liang}}\ and\ \bibinfo {author} {\bibfnamefont {X.~P.}\ \bibnamefont
  {Gao}},\ }\href {\doibase 10.1021/nl301325h} {\bibfield  {journal} {\bibinfo
  {journal} {Nano Letters}\ }\textbf {\bibinfo {volume} {12}},\ \bibinfo
  {pages} {3263} (\bibinfo {year} {2012})}\BibitemShut {NoStop}%
\bibitem [{\citenamefont {Alicea}(2012)}]{ali12}%
  \BibitemOpen
  \bibfield  {author} {\bibinfo {author} {\bibfnamefont {J.}~\bibnamefont
  {Alicea}},\ }\href@noop {} {\bibfield  {journal} {\bibinfo  {journal}
  {Reports on Progress in Physics}\ }\textbf {\bibinfo {volume} {75}},\
  \bibinfo {pages} {076501} (\bibinfo {year} {2012})}\BibitemShut {NoStop}%
\bibitem [{\citenamefont {Li}\ \emph {et~al.}(2016)\citenamefont {Li},
  \citenamefont {Neupert}, \citenamefont {Bernevig},\ and\ \citenamefont
  {Yazdani}}]{li16}%
  \BibitemOpen
  \bibfield  {author} {\bibinfo {author} {\bibfnamefont {J.}~\bibnamefont
  {Li}}, \bibinfo {author} {\bibfnamefont {T.}~\bibnamefont {Neupert}},
  \bibinfo {author} {\bibfnamefont {B.~A.}\ \bibnamefont {Bernevig}}, \ and\
  \bibinfo {author} {\bibfnamefont {A.}~\bibnamefont {Yazdani}},\ }\href
  {http://dx.doi.org/10.1038/ncomms10395} {\bibfield  {journal} {\bibinfo
  {journal} {Nature Communications}\ }\textbf {\bibinfo {volume} {7}},\
  \bibinfo {pages} {10395 EP } (\bibinfo {year} {2016})}\BibitemShut {NoStop}%
\bibitem [{\citenamefont {Saha}\ \emph {et~al.}(2014)\citenamefont {Saha},
  \citenamefont {Rainis}, \citenamefont {Tiwari},\ and\ \citenamefont
  {Loss}}]{sah14}%
  \BibitemOpen
  \bibfield  {author} {\bibinfo {author} {\bibfnamefont {A.}~\bibnamefont
  {Saha}}, \bibinfo {author} {\bibfnamefont {D.}~\bibnamefont {Rainis}},
  \bibinfo {author} {\bibfnamefont {R.~P.}\ \bibnamefont {Tiwari}}, \ and\
  \bibinfo {author} {\bibfnamefont {D.}~\bibnamefont {Loss}},\ }\href {\doibase
  10.1103/PhysRevB.90.035422} {\bibfield  {journal} {\bibinfo  {journal} {Phys.
  Rev. B}\ }\textbf {\bibinfo {volume} {90}},\ \bibinfo {pages} {035422}
  (\bibinfo {year} {2014})}\BibitemShut {NoStop}%
\bibitem [{\citenamefont {Rainis}\ \emph {et~al.}(2014)\citenamefont {Rainis},
  \citenamefont {Saha}, \citenamefont {Klinovaja}, \citenamefont {Trifunovic},\
  and\ \citenamefont {Loss}}]{rai14}%
  \BibitemOpen
  \bibfield  {author} {\bibinfo {author} {\bibfnamefont {D.}~\bibnamefont
  {Rainis}}, \bibinfo {author} {\bibfnamefont {A.}~\bibnamefont {Saha}},
  \bibinfo {author} {\bibfnamefont {J.}~\bibnamefont {Klinovaja}}, \bibinfo
  {author} {\bibfnamefont {L.}~\bibnamefont {Trifunovic}}, \ and\ \bibinfo
  {author} {\bibfnamefont {D.}~\bibnamefont {Loss}},\ }\href {\doibase
  10.1103/PhysRevLett.112.196803} {\bibfield  {journal} {\bibinfo  {journal}
  {Phys. Rev. Lett.}\ }\textbf {\bibinfo {volume} {112}},\ \bibinfo {pages}
  {196803} (\bibinfo {year} {2014})}\BibitemShut {NoStop}%
\bibitem [{\citenamefont {Chang}\ \emph {et~al.}(2014)\citenamefont {Chang},
  \citenamefont {van~den Brink},\ and\ \citenamefont {Ortix}}]{cha14}%
  \BibitemOpen
  \bibfield  {author} {\bibinfo {author} {\bibfnamefont {C.-H.}\ \bibnamefont
  {Chang}}, \bibinfo {author} {\bibfnamefont {J.}~\bibnamefont {van~den
  Brink}}, \ and\ \bibinfo {author} {\bibfnamefont {C.}~\bibnamefont {Ortix}},\
  }\href {\doibase 10.1103/PhysRevLett.113.227205} {\bibfield  {journal}
  {\bibinfo  {journal} {Phys. Rev. Lett.}\ }\textbf {\bibinfo {volume} {113}},\
  \bibinfo {pages} {227205} (\bibinfo {year} {2014})}\BibitemShut {NoStop}%
\bibitem [{\citenamefont {Ying}\ \emph {et~al.}(2016)\citenamefont {Ying},
  \citenamefont {Gentile}, \citenamefont {Ortix},\ and\ \citenamefont
  {Cuoco}}]{yin16}%
  \BibitemOpen
  \bibfield  {author} {\bibinfo {author} {\bibfnamefont {Z.-J.}\ \bibnamefont
  {Ying}}, \bibinfo {author} {\bibfnamefont {P.}~\bibnamefont {Gentile}},
  \bibinfo {author} {\bibfnamefont {C.}~\bibnamefont {Ortix}}, \ and\ \bibinfo
  {author} {\bibfnamefont {M.}~\bibnamefont {Cuoco}},\ }\href {\doibase
  10.1103/PhysRevB.94.081406} {\bibfield  {journal} {\bibinfo  {journal} {Phys.
  Rev. B}\ }\textbf {\bibinfo {volume} {94}},\ \bibinfo {pages} {081406}
  (\bibinfo {year} {2016})}\BibitemShut {NoStop}%
\bibitem [{\citenamefont {Chang}\ and\ \citenamefont {Ortix}(2017)}]{cha17}%
  \BibitemOpen
  \bibfield  {author} {\bibinfo {author} {\bibfnamefont {C.-H.}\ \bibnamefont
  {Chang}}\ and\ \bibinfo {author} {\bibfnamefont {C.}~\bibnamefont {Ortix}},\
  }\href {\doibase 10.1021/acs.nanolett.7b00426} {\bibfield  {journal}
  {\bibinfo  {journal} {Nano Letters}\ }\textbf {\bibinfo {volume} {17}},\
  \bibinfo {pages} {3076} (\bibinfo {year} {2017})}\BibitemShut {NoStop}%
\end{thebibliography}
\end{document}